\documentclass[aps,twocolumn,amsmath,amssymb,preprintnumbers]{revtex4}
\usepackage{amsmath} \usepackage{amsfonts} \usepackage{amssymb}
\usepackage{bbm}
\usepackage{epsfig}
\usepackage{graphics}
\usepackage{graphicx}
\newcommand{\be}{\begin{equation}}
\newcommand{\ee}{\end{equation}}
\newcommand{\ba}{\begin{eqnarray}}
\newcommand{\ea}{\end{eqnarray}}
\newcommand{\nn}{\nonumber}
\newcommand{\hmn}{_{\hat\mu\hat\nu}}

\newcommand{\mn}{_{\mu\nu}}
\newcommand{\iv}{^{(4)}}
\newcommand{\iD}{^{(D)}}
\newcommand{\ib}{^{1/2}}
\newcommand{\hmnrs}{_{\hat\mu\hat\nu\hat\rho\hat\sigma}}
\newcommand{\ihmnrs}{^{\hat\mu\hat\nu\hat\rho\hat\sigma}}
\newcommand{\mnrs}{_{\mu\nu\rho\sigma}}

\begin{document}

\title[ ]{Dilatation symmetry in higher dimensions and the vanishing of the cosmological constant}

\author{C. Wetterich}
\affiliation{Institut  f\"ur Theoretische Physik\\
Universit\"at Heidelberg\\
Philosophenweg 16, D-69120 Heidelberg}

\begin{abstract}
A wide class of dilatation symmetric effective actions in higher dimensions leads to a vanishing four-dimensional cosmological constant. This requires no tuning of parameters and results from the absence of an allowed potential for the scalar dilaton field. The field equations admit many solutions with flat four-dimensional space and non-vanishing gauge couplings. In a more general setting, these are candidates for asymptotic states of cosmological runaway solutions, where dilatation symmetry is realized dynamically if a fixed point is approached as time goes to infinity. Dilatation anomalies during the runaway can lift the degeneracy of solutions and lead to an observable dynamical dark energy.
\end{abstract}

\maketitle

Dilatation symmetry may play a crucial role in our understanding of cosmology \cite{CWQ}. If the quantum effective action is scale invariant, a spontaneous breaking of the dilatation symmetry by a non-zero value of the scalar dilaton field will lead to a massless Goldstone boson. In presence of a dilatation anomaly, a potential and a small mass for this field are generated - the pseudo-Goldstone boson becomes the cosmologically relevant ``cosmon field''. For many cosmological ``runaway solutions'' the dilatation anomaly vanishes asymptotically, as a fixed point is approached for time going to infinity \cite{CWCC}. In consequence, the mass of the cosmon field decreases with time. Typically, it is of the order of the Hubble parameter \cite{CWAA}. If the fixed point occurs for a vanishing cosmological constant, rather interesting quintessence cosmologies follow from this scenario \cite{CWQ}. They describe tracker solutions with dynamical dark energy \cite{CWQ,CWCC,CWAA,PR}. Our approach aims for an explanation why the cosmon potential vanishes asymptotically, rather than approaching a nonvanishing constant. 

In four dimensions, dilatation symmetry alone cannot explain why the fixed point occurs for a vanishing cosmological constant. Within a dilatation symmetric standard model of elementary particle physics, every mass scale is replaced by a combination  $h\chi$, with $\chi$ the scalar dilaton field and $h$ some appropriate dimensionless coupling. Four-dimensional dilatation symmetry permits a polynomial potential $V(\chi)=\lambda\chi^4$. After Weyl scaling, this results in the Einstein frame as an effective cosmological constant proportional to the dimensionless coupling $\lambda$. In contrast, dilatation symmetry in dimension $d>6$ does not allow anymore a polynomial potential. We will show in this note that for a wide class of dilatation symmetric effective actions this simple fact results in a vanishing four-dimensional cosmological constant without tuning of parameters. 

Consider the quantum effective action $\Gamma$ for the metric $\hat g\hmn$ and a dilaton field $\xi$ in arbitrary dimension $d$. It includes all effects from quantum fluctuations, including those of effective four-dimensional fields after spontaneous compactification. The field equations or other constraints \cite{CWCC} that arise from the extremum condition for $\Gamma$ are exact, all quantum effects being already included in the computation of $\Gamma$. Scale transformations (dilatations) act on the fields by multiplicative rescaling with powers of a constant $\nu,~\hat g\hmn\to\nu^{-2}\hat g\hmn~,~\xi\to\nu^{\frac{d-2}{2}}\xi$, while the coordinates $\hat x^{\hat\mu}$ remain unchanged. We will first explore the consequences of two simple assumptions: (i)the effective action is scale invariant; (ii)$\Gamma$ can be written as a polynomial of $\xi$ and the curvature scalar $\hat R_{\hat\mu\hat\nu\hat\rho\hat\sigma}$ or their covariant derivatives. 
We emphasize that we do {\em not} assume that the effective action as a whole has these properties. It is sufficient that our assumptions hold for the fixed point when the dilatation symmetry violating terms have vanished for a runaway solution approaching the fixed point arbitrarily close.

With our assumptions the most general dilatation symmetric effective action reads, with $\hat g=\det(-\hat g\hmn)$,
\be\label{2}
\Gamma=\int_{\hat x}\hat g^{1/2}\left\{-\frac12\xi^2\hat R+\frac\zeta2\partial^{\hat \mu}\xi\partial_{\hat\mu}\xi+F\right\}.
\ee
The first two terms are the higher dimensional generalization of the Jordan-Brans-Dicke theory \cite{JBD} in the absence of matter, while $F$ contains higher powers of the curvature tensor. The most striking property of \eqref{2} is the absence of a potential for $\xi$. Indeed, a scale invariant polynomial $\xi^n,n\in{\mathbbm N}$, requires $n(d-2)/2=d$. For $d=4$ one has $n=4$, for $d=6$ a cubic potential $\sim\xi^3$ is possible, but no solution exists for $d>6$. We will later weaken our assumptions and no longer require a polynomial form of $F$. 

The field equations derived from $\Gamma$ read
\ba\label{3}
&&\zeta D^{\hat \mu}D_{\hat\mu}\xi+\hat R\xi=0,\\
&&\xi^2(\hat R\hmn-\frac12\hat R\hat g\hmn)=T^{(\xi)}\hmn+T^{(F)}\hmn,\label{4}
\ea
with $T^{(\xi)}\hmn$ involving derivatives of $\xi$. For geometries with singularities the extremum condition for $\Gamma$ yields further ``brane constraints'' \cite{CWCB}. For example, an extremum with respect to the infinitesimal variation $\xi\to\xi\big(1+\epsilon(x)\big)$ requires a vanishing boundary term
\be\label{XA}
\int_y\partial_{\hat\mu}(\hat g^{1/2}\xi\partial^{\hat\mu}\xi)=0,
\ee
with $\int_y$ indicating the integration over the "internal coordinates'' $y^\alpha$, while $x^\mu$ denotes the four-dimensional coordinates. Similar brane constraints for the metric will be discussed below.

Neglecting first the contribution $T^{(F)}\hmn$ from $F$, the field equations have a simple class of static solutions, namely
\be\label{5}
\xi=\xi_0=const.\quad,\quad\hat R\hmn=0.
\ee
These are candidates for the asymptotic limits of time dependent cosmological solutions as $t\to\infty$. The condition of Ricci-flatness, $\hat R\hmn=0$, still allows many different geometries. Among them is $d$-dimensional Minkowski space. More interesting for the real world are geometries which are a direct product of flat four-dimensional Minkowski space and a Ricci-flat $D$-dimensional internal space $(D=d-4)$ with finite volume $\Omega_D$. Such geometries admit dimensional reduction to an effective four-dimensional theory. If we keep only the four-dimensional metric $g^{(4)}\mn(x)$ and a scalar dilaton field $\chi(x)=\Omega^{1/2}_D\xi(x)$, the reduced four-dimensional effective action $\Gamma_4$ exhibits an effective four dimensional dilatation symmetry
\be\label{7}
\Gamma^{(4)}=\int_x(g^{(4)})^{1/2}\left\{-\frac12\chi^2R^{(4)}+\frac{\zeta}{2}\partial^\mu\chi\partial_\mu\chi\right\}.
\ee
No term $\sim \lambda\chi^4$ appears - the effective four-dimensional cosmological constant vanishes. 

If internal space has isometries, $\Gamma^{(4)}$ can be extended to include the gauge bosons of the corresponding local gauge symmertry. For finite $\Omega_D$ the dimensionless gauge couplings are finite and non-zero. A simple example for a Ricci-flat internal space is a $D$-dimensional torus with isometry $U(1)^D$, but there are many more possibilities, including spaces with non-abelian isometries. Beyond a polynomial approximation to the effective action, the four-dimensional gauge coupling may be running, according to a gauge and dilatation invariant kinetic term $\sim F^{\mu\nu}K(-D^\mu D_\mu/\chi^2)F\mn$. For asymptotically free theories this can result in particle masses proportional to a ``confinement scale'' $\Lambda_c$ which scales $\sim\chi$. Ratios of particle masses and the effective Planck mass $\chi$ do not depend on $\chi$ in this case. Solutions \eqref{5} with finite $\Omega_D$ share therefore many aspects of a satisfactory asymptotic state of cosmology - namely flat space and particle physics with constant dimensionless couplings and mass ratios.

In even dimensions we may include a suitable polynomial $F$ of the curvature tensor $\hat R_{\hat\mu\hat\nu\hat\rho\hat\sigma}$ and its covariant derivatives. An example is $F=\tau\hat R^{\frac d2}.$ Scale invariance requires $\frac d2-m$ powers of $\hat R_{\hat\mu\hat\nu\hat\rho\hat\sigma}$, with $m$ the number of covariant derivatives. Indices have to be contracted with $\hat g^{\hat\mu\hat\nu}$, such that $m$ must be even. (We disregard here parity violating contractions with the $\epsilon$-tensor.) This excludes all possible polynomials $F$ for $d$ odd, such that our assumptions imply $F=0$ for this case. In principle, there may be also polynomials with one power of $\xi$ and $\frac d4+\frac 12-m$ powers of $\hat R_{\hat\mu\hat\nu\hat\rho\hat\sigma}$. This is only possible for $d=2~mod~4$. We may exclude such invariants by a discrete symmetry $\xi\to -\xi$. We disregard a total derivative $(\hat D^2)^{(d-2)/2}\hat R$ - therefore $F$ contains at least two powers of the curvature tensor. 

Consider first the case where $F$ can be written as a polynomial of $\hat R\hmn$ and its covariant derivatives. It is easy to see that the solution \eqref{5} solves the field equations derived from this large class of dilatation symmetric effective actions. The variation of $F$ always contains terms linear in $\hat R\hmn$ and its covariant derivatives and therefore gives no contribution if $\hat R\hmn=0$. This general form of $F$ contains a large number of different invariants, with different dimensionless couplings $\tau_i$. We have therefore established solutions for which the effective four-dimensional cosmological constant vanishes for arbitrary $\tau_i$, involving no tuning of parameters. We can generalize this argument for the most general form of $F$ which can be written as a polynomial of $\hat R$, the traceless part of the Ricci tensor $\hat H\hmn=\hat R\mn-\frac1d\hat R\hat g\hmn$, and the totally antisymmetric part of the curvature tensor $\hat C\hmnrs=\hat R_{[\hat\mu\hat\nu\hat\rho\hat\sigma]}$, as well as covariant derivatives thereof. Flat space geometries, $\hat R\hmnrs=0,\xi=\xi_0$, always solve the field equations. They include the direct product of four-dimensional Minkowski space and a $D$-dimensional torus.

Beyond flat space, there are many other possibilities of extrema of $\Gamma$ with $\Lambda=0$. We may write $F=F_2+F_C$, where $F_2$ contains two or more powers of $\hat R\hmn$. Solutions with $\hat C\hmnrs\neq 0,~\hat R\hmn=0,\Lambda=0$ require $\int_y\hat g\ib F_C=0$. One can give examples where this condition cannot be met, as $F_C=\tau_c(\hat C\hmnrs C\ihmnrs)^{d/4}$. For $\tau_c\neq 0$ one may rather have extrema with $\Lambda=0$ and both $\hat H\hmn$ and $\hat C\hmnrs$ vanishing or both non-vanishing. Furthermore, a constant $\xi=\xi_0$ implies $\hat R=0$ by virtue of the scalar field equation \eqref{3}, while extrema with $\hat R \neq 0$ remain consistent with $\partial^\alpha\xi\partial_\alpha\xi\neq 0$.

Our assumptions imply the presence of asymptotic solutions with a vanishing effective four dimensional cosmological constant already under rather general conditions. We next turn to a more general discussion of ``quasistatic'' geometries, for which the internal geometry and $\xi$ are independent of $x^\mu$, while the four-dimensional space-time is maximally symmetric. The four-dimensional Ricci tensor obeys $R\iv\mn=\Lambda g\iv\mn$, where positive (negative) $\Lambda$ corresponds to (anti-) de Sitter space, and a vanishing cosmological constant, $\Lambda=0$, to flat Minkowski space. For this discussion we weaken our assumptions and require no longer a polynomial dependence of $F$ on the curvature tensor. For this very general setting we find as a central result of this note that all stable extrema of the effective action with nonzero $\xi$ and finite characteristic length scale $l$ of internal space have a vanishing cosmological constant. More precisely, we can show that for $|\Lambda|\ll\chi^2$ no stable extrema of $\Gamma$ with $\Lambda\neq 0$ exist in this case. Stable quasistatic solutions single out a vanishing cosmological constant!

The direct product solutions are not the only interesting geometries. There may be singular solutions with warping \cite{RSW,CWWB,RDW,RS}, corresponding to a brane \cite{ARS,CWWB} or a ``zerowarp'' \cite{RSW} sitting at the singularity. Such spaces may be interesting because they can lead to chiral fermions after dimensional reduction \cite{CWCF}. The most general quasistatic geometry 
\be\label{8A}
\hat g\hmn=\left(\begin{array}{ccc}
\sigma(y)g\iv\mn(x)&,&0\\
0&,&g\iD_{\alpha\beta}(y)
\end{array}
\right),
\ee
involves the warp factor $\sigma(y)$ and the internal metric $g_{\alpha\beta}(y)$. For singular solutions one can always obtain local solutions with $\Lambda=0$, but often also neighboring solutions with $\Lambda\neq 0$ exist \cite{RSW,CWWB,RDW}. In the absence of additional physics that fixes the strength of the singularity, we find that some of the solutions with $\Lambda=0$ (flat four-dimensional geometry) are consistent extrema of the action \cite{CWCB}. In contrast, neighboring solutions with a small $|\Lambda|\ll \chi^2$ are inconsistent and do not correspond to extrema of $\Gamma$. The remarkable property that a vanishing cosmological constant is singled out originates from the observation that for unconstrained higher-dimensional fields (``variable tension branes'' in case of singularities \cite{CWCB}) - the extremum conditions go beyond the higher-dimensional field equations. Additional ``brane constraints'' \cite{CWCB} are due to eq. \eqref{XA} and to similar boundary terms for the metric. For regular geometries the brane constraints are obeyed automatically. 

We restrict our discussion to solutions with local four-dimensional gravity, in the sense that an effective four-dimensional action can be expanded in powers of the curvature tensor. In this case an extremum of the higher-dimensional action must also be an extremum of the effective four-dimensional action. The latter obtains from $\Gamma$, eq. \eqref{2}, by inserting given solutions of the higher dimensional field equations $\xi(y),\sigma(y)$ and $g^{(D)}_{\alpha\beta}(y)$ according to the ansatz \eqref{8A}, and integrating over the internal coordinates $y$. In particular, we can consider $\Gamma\iv$ as a functional of the variable $g\iv\mn$
\be\label{X1}
\Gamma\iv=\int_x(g\iv)^{1/2}W~,~W=V-\frac{\chi^2}{2}R\iv+\tilde H(R\iv\mnrs).
\ee
Here we identify $V$ with the effective four-dimensional cosmological constant, $\chi$ with the effective Planck mass (in the Jordan frame) and $\tilde H$ contains higher orders of the four-dimensional curvature tensor. For a maximally symmetric four-dimensional space one finds for the four-dimensional Lagrangian $W$
\be\label{X2}
W=V-2\Lambda\chi^2+\Lambda^2\hat H(\Lambda)~,~\lim_{\Lambda\to 0}\Lambda\hat H(\Lambda)=0.
\ee
Both $V$ and $\chi$ depend on the geometry of internal space and on the warping.

In addition to $g\iv\mn$ we may also keep the normalization of $\xi$ and $g\iD_{\alpha\beta}$ as free variables. These degrees of freedom can be expressed in terms of a characteristic length scale $l$ for internal space and a characteristic average $\bar\xi$ for $\xi$
\be\label{X3}
\int_y(g\iD)^{1/2}\sigma^2=l^D~,~\int_y(g\iD)\ib\sigma\xi^2=l^D\bar\xi^2.
\ee
We are interested in extrema where both $l$ and $\bar\xi$ are finite and non-zero. For the metric \eqref{8A} the $d$-dimensional curvature scalar obeys $\hat R=R^{(\textup{int})}+R\iv/\sigma$ such that
\be\label{X4}
\chi^2=l^D\bar\xi^2-2\tilde Gl^{-2}~,~\tilde G=l^2\int_y(g\iD)\ib\sigma G.
\ee
Here $G$ arises from the expansion of $F$ in the four-dimensional curvature tensor $F(\hat R\hmnrs)=F(R^{(\textup{int})}\hmnrs)+GR\iv/\sigma+\dots$, and $\tilde G$ is dimensionless. Similarly, we can write
\be\label{X6}
V=\tilde Q\bar\xi^2l^{D-2}+\tilde F l^{-4},
\ee
with dimensionless quantities
\ba\label{X7}
\tilde Q&=&\frac12\bar\xi^{-2}l^{2-D}\int_y(g\iD)\ib\sigma^2
(\zeta\partial^\alpha\xi\partial_\alpha\xi-
\xi^2 R^{(\textup{int})}),\nn\\
\tilde F&=&l^4\int_y(g\iD)\ib\sigma^2F(R^{(\textup{int})}\hmnrs).
\ea
Eq. \eqref{X2} yields for arbitrary quasistatic geometries
\be\label{X8}
W=\bar\xi^2\left[\tilde Q l^{D-2}-2\Lambda l^D\right]
+\frac{\tilde F} {l^4}+\frac{4\tilde G\Lambda}{l^2}+\Lambda^2\hat H(\Lambda l^2).
\ee

An extremum of $\Gamma$ must be an extremum of $W(\bar\xi,l)$, provided that $g\iv\mn$ and therefore $\Lambda$ is kept fixed. Stability requires $\tilde Q\geq 0~,~\tilde F\geq 0~,~\chi^2>0$ and the absence of negative eigenvalues for the second variations of $W$. For $|\Lambda|\ll\chi^2$ we can further neglect the term $\Lambda^2\hat H$. A general analysis of the stable extrema of $W$ shows two ``phases''. A ``flat phase'' with $\Lambda=0$ is possible for all $\tilde F,\tilde Q,\tilde G$. For $\tilde F>0$ it occurs, however, for $l\to\infty$. A ``non-flat phase'' with $\Lambda\neq 0$ is possible only in a restricted range of $(\tilde F,\tilde Q,\tilde G)$ with $\tilde G<0,\tilde Q>0$, and requires $\bar\xi=0$. All stable extrema with $\bar\xi>0$ belong to the flat phase with $\Lambda=0$. Furthermore, a finite volume $l<\infty$ requires $\tilde F=\tilde Q=0$. The solutions in the flat phase remain intact if we add the term $\Lambda^2\hat H$. It is conceivable that in presence of $\hat H\neq 0$ new stable extrema become possible with $|\Lambda|\approx \chi^2$. In units of the scale $l^D\bar \xi^2$ the possible values of $\Lambda$ show a gap between zero and a value of order unity. In view of the discreteness characteristic for the extrema of $\Gamma$ \cite{CWCB} and the existence of solutions with $\Lambda=0$, this may not be  too surprising. 

The phase structure of the possible stable extrema of $\Gamma$ has important consequences for the stability of $\Lambda$ with respect to parameter changes of the higher dimensional action. Let us start with a specific form of $\Gamma$ for which solutions in the flat phase with $\xi>0,l<\infty$ are explicitly known. An example is a polynomial $F$ with $F_C=0$ and direct product solutions of four dimensional Minkowski space and Ricci flat internal space, for which $\Lambda=0,\tilde F=\tilde Q=0$ can be easily verified. We next modify $\Gamma$ by a small change $\delta F$ of the $\xi$-independent term. Either no solution with $\xi>0,l<\infty$ exists anymore in the presence of $\delta F$. This could happen if $\delta F$ induces an instability for all such solutions. Or the modified $\Gamma$ still admits stable extrema with $\xi>0,l<\infty$. Then the flat phase persists and $\Lambda=0$ remains preserved - a jump to a value $\Lambda\approx \chi^2$ is not possible for a continuous change of $F$ (except for special points where $\chi^2$ vanishes). We may evaluate the contribution of $\delta F$ to $\tilde F$, i.e. $\delta \tilde F$, for the original extremum of $\Gamma$ (i.e. in the absence of $\delta F$). It is not necessary that $\delta\tilde F$ vanishes. If not, the internal geometry and $\sigma(y),\xi(y)$ will readjust such that for the new extremum in the presence of the term $\delta F$ one again finds $\tilde F=0$. This produre can be continued to ``explore'' the parameter space of effective actions for which stable extrema with $\xi>0,l<\infty$ exist - they all have $\Lambda=0$. Typically this parameter space covers arbitrary $F$ which are consistent with general stability criteria, since flat internal space with finite $\Omega_D$ and $\xi=\xi_0$ always provides for a possible extremum with $\xi>0,l<\infty$. 

The readjustment of $\tilde F$ to zero can be associated with a readjustment of the cosmological constant. (Note that the readjustment of $\tilde Q$ follows automatically, since for any configuration \eqref{8A} with $g\iv\mn=\eta\mn$ the field equation \eqref{3} for $\xi$ and the extremum condition \eqref{XA} imply $\tilde Q=0$ \cite{CWCB}.) In this respect we emphasize an important difference between higher dimensional theories and the four dimensional setting. In general, $\tilde F$ is a functional of higher dimensional fields $\alpha(y)$ that describe the changes of $g\iD_{\alpha\beta},\sigma$ and $\xi$. The extremum condition $\delta\tilde F/\delta\alpha(y)$ amounts to field equations with derivatives of $\alpha$. Local solutions have free integration constants which may be adapted to obey $\tilde F=0$. There is no need for a very special form of $\tilde F[\alpha(y)]$. In other words, we have infinitely many four dimensional fields in order to achieve the readjustment. This contrasts with a finite number of homogeneous four dimensional fields $\alpha_i$, where the conditions $\partial\tilde F/\partial\alpha_i=0,\tilde F=0$ can be met simultaneously only for a special choice of $\tilde F$. (This is one aspect of the ``fine tuning problem'' for the cosmological constant.)

These arguments equally apply for the role of particular ``Casimir-contributions'' to $\Lambda$ from some effective four dimensional quantum fluctuations, or  from the QCD-condensate $\Lambda_{QCD}$ or the Fermi scale. In a dilatation symmetric setting these effects typically are $\sim l^{-4}$ and therefore give a contribution to $\tilde F$. In a general covariant setting they therefore contribute to $F$, perhaps in a non-polynomial form. Indeed,  the ``compactification scale'' $l^{-1}$ acts as an effective ultraviolet cutoff for the validity of a four dimensional description and sets the scale for Casimir effects. It also sets the initial scale for the four dimensional running of couplings, such that $\Lambda_{QCD}\sim l^{-1}$ etc.. Dilatation symmetry implies that the values of dimensionless couplings evaluated at $l^{-1}$ can only depend on dimensionless combinations of scales as $\omega=\bar\xi l^{(D+2)/2}$. (In this respect we assume that a small possible residual non quadratic $\bar\xi$-dependence of $\Gamma$ form a possible $\xi$-dependence of $\Lambda_{QCD}$ does not alter our analysis, at least in the limit $\bar\xi\to\infty,\omega\to$ const in which we are interested finally, see below.)

We have seen that there are many other contributions to $\tilde F$ from geometrical degrees of freedom. If a stable extremum with $\bar\xi>0,l<\infty$ exists, all these contributions must cancel precisely by virtue of the higher dimensional field equations, resulting in $\tilde F=0$. Tiny adjustments of infinitely many four dimensional scalar fields are sufficient for this purpose. Typically, such a field $A$ has a mass term $\sim \chi^2A^2$ and couples linearly to the low energy degrees of freedom, as $\chi\varphi^\dagger \varphi A$ for the coupling to the Higgs field $\varphi$. A change of $\varphi\to\varphi+\delta\varphi$ results in a change of $V=\Lambda\chi^2$ of the order $\delta V\sim\varphi^3\delta\varphi$ due to quartic interactions $\lambda\varphi^4$ etc.. It is easy to verify that the resulting change $\delta A\sim \varphi\delta\varphi/\chi$ also contributes $\delta V\sim \varphi^3\delta\varphi$, allowing for compensation. 

Dilatation symmetry is not expected to be an exact quantum symmetry. Even for a scale invariant classical or microscopic action, the quantum fluctuations typically induce a dilatation anomaly. This originates from the missing dilatation invariance of the measure in the functional integral. Nevertheless, a dilatation symmetric effective action can play a particular role in cosmology. This is due to the possibility of cosmological ``runaway solutions'', where fields change in time over all the cosmological history, including the present epoch, such that $\Gamma$ approaches effectively a dilatation symmetric fixed point.

The issue of the dilatation anomaly can be understood by performing a Weyl scaling of the $d$-dimensional metric 
$\hat g\hmn=w^2\tilde g\hmn~,~w=M_d\xi^{-\frac{2}{d-2}}$. With $\tilde R$ the curvature scalar computed from the metric $\tilde g\hmn$ in the Einstein frame, the effective action \eqref{2} reads in the new fields
\ba\label{X20}
\Gamma&=&\int\tilde g\ib\big\{-\frac{M^{d-2}_d}{2}
(\tilde R-\tilde \zeta\partial^{\hat\mu}\ln\xi\partial_{\hat\mu}\ln\xi)\nn\\
&&+\tilde F(\tilde R\hmnrs,\partial_{\hat\mu}\ln \xi)\big\},
\ea
where $\ln\xi$ stands for $\ln(\xi/M^{(d-2)/2}_d)$. In the Einstein frame \eqref{X20} the dilatations act as shifts in the dilaton field $\delta\sim M_d\ln\xi$, while $\tilde g\hmn$ is invariant. Quantization in the Einstein frame preserves the dilatation symmetry as an exact quantum symmetry, since a functional measure for $\delta$ is invariant under a global shift $\delta\to\delta+\alpha$. However, we propose that the functional measure is defined in terms of the original variables $\xi$ and $\hat g\hmn$. The Weyl scaling to a measure for $\delta$ and $\tilde g\hmn$ involves a Jacobian $\sim w^f$ for every spacetime point $\hat x$, and therefore contributes to $\Gamma$ an anomalous piece $\Gamma_{an}\sim\Sigma_{\hat x}\ln\xi^2$. Regularization of the sum over all spacetime points introduces a mass scale $\mu$ which explicitly breaks dilatation symmetry, $\Gamma_{an}=\int_{\hat x}\hat g^{1/2} \mu^d(\ln\xi^2+$const.), where $\hat g^{1/2}$ arises from the requirement of general covariance of the regularization. 

To demonstrate the effect of the anomaly we add in the bracket in eq. \eqref{2} a term $\hat V_{an}=\mu^d$. After dimensional reduction the anomaly adds to $V$ in eq. \eqref{X1}  a term $V_{an}=\mu^dl^D$. In order to discuss cosmological solutions it is convenient to perform a four dimensional Weyl scaling with $w_4=M/\chi$, such that in terms of the new metric for the Einstein frame one has $\Gamma\iv=\int g^{1/2}(-M^2 R/2+U)$. For $\tilde F=\tilde Q=0$ the effective potential reads
\be\label{ZA}
U=\frac{M^4V_{an}}{\chi^4}=M^4\gamma^{\frac{D}{2}}\left(\frac{\mu}{\chi}\right)^d,
\ee
where we have introduced $\gamma=\chi^2l^2=\omega^2-2\tilde G$. Realistic asymptotic solutions should lead to a constant value $\omega$. Cosmology corresponds then to an increase of $\chi$ for $t\to\infty$, resulting in a decrease of the effective cosmological constant $\sim U/M^2$. This yields a typical quintessence cosmology, with an exponentially decreasing potential for the cosmon field $\varphi\sim\ln(\chi/M)$ \cite{CWQ}. The effect of the anomaly vanishes for $t\to\infty$ such that the system tends indeed to one of the quasistatic extrema of a dilatation symmetric effective action. 

We conclude that a cosmological runaway towards a fixed point, where dilatation symmetry is realized dynamically on a quantum level, offers exciting prospects for a solution of the cosmological constant problem. The asymptotic value of the cosmological constant vanishes for a wide class of dilatation symmetric asymptotic states, without any tuning of parameters.

\end{document}